# Differential optical transfer function wavefront sensing

Johanan L. Codona

# Differential optical transfer function wavefront sensing

**Johanan L. Codona**
University of Arizona
Center for Astronomical Adaptive Optics
Steward Observatory
Tucson, Arizona 85721
E-mail: jlcodona@gmail.com

**Abstract.** An image-based technique for measuring the complex field in the pupil of an imaging system is presented. Two point source images, one with a small modification introduced in the pupil, are combined using a simple and non-iterative algorithm. The non-interferometric method is based on the change in the optical transfer function (OTF) giving a differential optical transfer function (dOTF). The dOTF includes two images of the complex pupil field, conjugated and reflected about the position of the pupil modification, leaving an overlap that obscures some of the pupil. The overlap can be minimized by introducing the modification near the edge of the pupil. The overlap region can be eliminated altogether by using a second modification and a third point source image. The pupil field is convolved by the change in the pupil field, so smaller modification areas are preferred. When using non-monochromatic light, the dOTF incurs a proportional radial blurring determined by the fractional bandwidth. We include some simple demonstration experiments, including using a pupil blockage and moving a single deformable mirror actuator as the pupil modification. In each case, the complex wavefront is easily recovered, even when the pupil mask is unknown and the wavefront aberrations are large. © *The Authors. Published by SPIE under a Creative Commons Attribution 3.0 Unported License. Distribution or reproduction of this work in whole or in part requires full attribution of the original publication, including its DOI.* [DOI: 10.1117/1.OE.52.9.097105]

Subject terms: wavefront sensors; optical transfer functions; point spread functions; optical testing.

Paper 131045P received Jul. 12, 2013; revised manuscript received Aug. 13, 2013; accepted for publication Aug. 13, 2013; published online Sep. 20, 2013.

## 1 Introduction

In a system where the point spread function (PSF) is shifted invariantly across the field, the incoherent object and image intensities are related by convolution with the PSF. The optical transfer function (OTF) allows this same operation to be expressed as a multiplication in the Fourier domain. The OTF is the inverse Fourier transform of the PSF, or equivalently, the average of the mutual coherence function (MCF) over the pupil plane for a given vector baseline.[1] As the spatial filter affecting the optical resolution of an imaging system, the OTF as well as its amplitude [the modulation transfer function (MTF) = abs(OTF)] and phase [the phase transfer function (PTF) = −arg(OTF)] are very familiar and well studied.[1] The OTF is inherently non-linear because the PSF is a quadratic functional of the pupil field. The OTF and its constituent functions are often used to characterize shortcomings of an optical system, although they would be more easily diagnosed if the aberrated or vignetted pupil field were known instead of just the OTF. It is a matter of knowing cause versus effect.

Estimating the pupil wavefront is often approached more directly, with measurements based on the addition of some other wavefront sensor (WFS) optics, such as a Shack–Hartmann sensor.[2] These methods also detect intensity, but over sub-apertures or some other field partitioning that can be processed post-detection to give an estimate of the wavefront in the pupil plane. Some wavefront sensing methods are very photon efficient and give reasonably accurate measurements at low flux levels, enabling high-speed applications such as adaptive optics (AO). Any separate WFS approach that employs extra optics or uses diverted beams and extra sensors to estimate the field, runs the risk of non-common-path (NCP) errors where aberrations or vignetting affecting either the final image or the WFS estimates are introduced on only one branch of the optical system. A familiar problem is when an AO system's WFS reports that the wavefront has been flattened to some accuracy, while the final image may still be distorted due to aberrations introduced downstream from the WFS. In this case, we are left to find the cause and remedy by other means. The only way to ensure avoidance of NCP errors is to make the wavefront measurements with the intended imaging sensor in the intended configuration.

Since the OTF is the pupil field convolved with its own complex conjugate, it is tantalizing to imagine retrieving the pupil field from it, or by extension, from an image of the PSF. But even though it is easy to acquire an image of a star or a laser spot in the laboratory, techniques such as the "star test"[3] are not simple enough to interpret beyond classic low-order aberrations, and even then the signs of the aberrations can be lost or seriously confused. More sophisticated methods have been developed for determining the pupil field (phase and amplitude) from PSFs and a knowledge of the pupil configuration. These include iterative transform algorithms such as Gerchberg–Saxton and other model-based algorithms that attempt to find the phase of the pupil field from the PSF and various constraints using gradient search techniques.[4,5] Adding more information, such as one or more defocused (or otherwise phase diverse) images,[6] helps to avoid some of the degeneracies of the basic problem. Images taken with the system in two or more known configurations allow an iterative solution to be found. These techniques can be very powerful when applicable, but they remain challenging and prone to failure if various assumptions (such as pixel

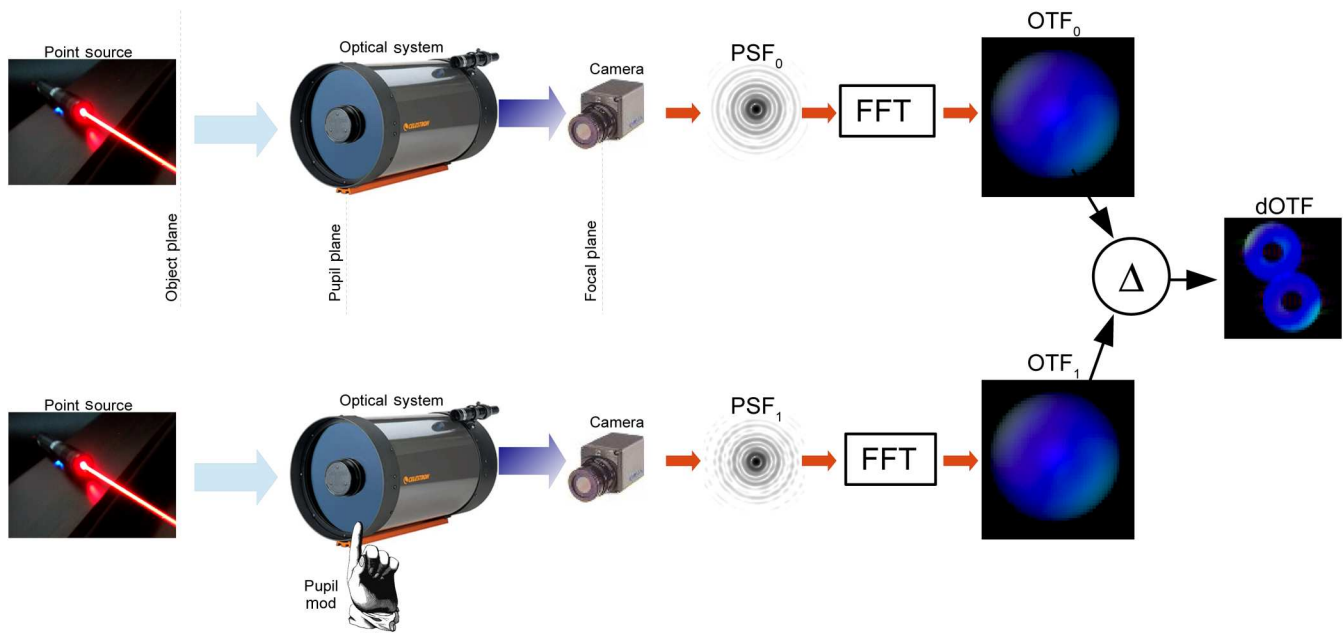

**Fig. 1** The process of acquiring a dOTF measurement. An optical system under test is used to image a point source. This can be an artificial source in the laboratory or a band-passed star. A camera is used to capture the image of the point source, which is then Fourier transformed to give the complex OTF. The same point source is imaged again with a pupil modification inserted somewhere near the edge. The figure shows a finger being used for this purpose, which is actually a feasible option. Since the dOTF does not need to be calibrated and is essentially self-documenting as to the modification's location and size, inserting a finger is actually a perfectly viable method for making a highly accurate pupil field measurement. Once the two PSFs are acquired and Fourier transformed, the dOTF is found by subtracting one from the other. If the source flux varied between the images, they may be relatively normalized by scaling one until the resulting dOTF is dark outside the pupil image and its reflection.

size in $\lambda/D$ or the vignetted shape or transmission profile of the pupil) are incorrect.

This article presents a new technique for measuring the complex field in the pupil plane from the difference between the Fourier transform of two PSF images [differential optical transfer function (dOTF)]. The technique is image-based, non-interferometric, and the processing is non-iterative. The motivation for dOTF is the simple idea that a quadratic form is made linear by differentiation. We explore this idea and develop it into a simple and effective method for determining the complex field over the majority of the pupil (Fig. 1). The two required images are of a single point source, where one is taken with the desired pupil and the second is taken with the pupil altered over some small region, such as by blocking a portion of it near its edge or poking a single outlying actuator with a deformable mirror (DM). The difference between the Fourier transforms of these images approximates the functional derivative[7,8] of the OTF with respect to the pupil mask, giving a single "term" in the MCF integration comprising the OTF. Due to the symmetries involved in computing the focal plane intensity from the field, the functional derivative of the OTF leaves us with two redundant overlapping pupil copies complex conjugated and reflected about the point of pupil modification.[8] Where the two images do not overlap, the pupil field stands alone and is able to be examined in both phase and amplitude. In the overlapping region, the portion of the pupil field that is anti-Hermitian (anti-symmetric real and symmetric imaginary parts) about the modification point is obscured. But, if the pupil modification is introduced near the edge of the pupil, the overlapping region is small. If a third PSF is acquired with a second pupil modification near some other part of the pupil's edge, the new overlap region will be located at a different place and the pair of dOTF images can be combined to retrieve the field at all points across the pupil. The dOTF method is extremely simple to implement, and if the images are taken with the actual image sensor, the problem of NCP aberrations is eliminated. The result gives the phase and amplitude in the pupil plane as seen from the final focal plane, including all effects such as aberrations, misalignments, and vignetting, projected back to the pupil. Depending on the nature of the pupil modification, measurements are simple to make and the result can be interpreted directly without detailed calibration.

In Sec. 2, we see how the functional derivative of the OTF gives rise to this simple result and how more realistic images with larger and more distributed pupil modifications simply blur the estimate. The dOTF theory is put to a numerical test in Sec. 2.3 with a simple Fourier optics calculation of a telescope observing a star through a phase screen with a small opaque pupil blockage near its edge. In Sec. 2.5, we consider some other implementations, including a single actuator poke using a DM. In Sec. 3, we perform some initial experimental demonstrations to validate the dOTF theory. Finally, in Sec. 4, we consider some applications of the method.

## 2 Theory

We begin this discussion using monochromatic Fourier optics[9] at wavelength $\lambda$ ($k = 2\pi/\lambda$). The pupil field is the product of the complex incident field $\psi_0(\mathbf{x}) = \alpha(\mathbf{x}) \exp[i\phi(\mathbf{x})]$ and a (possibly complex) pupil transmission mask

$$\psi(\mathbf{x}) = \Pi(\mathbf{x})\psi_0(\mathbf{x}), \qquad (1)$$

where $\mathbf{x}$ is a two-dimensional (2-D) transverse vector in the pupil plane. The incident field $\psi_0(\mathbf{x})$ may arise from any coherent monochromatic source distribution in the object plane, but we will assume that it arises from a "point source" for descriptive convenience. The pupil field is the superposition of plane waves with transverse spatial frequencies $\boldsymbol{\kappa} = k\boldsymbol{\theta}$ and complex amplitudes

$$\Psi(\boldsymbol{\kappa}) = \int e^{i\boldsymbol{\kappa}\cdot\mathbf{x}}\psi(\mathbf{x})\mathrm{d}^2x. \quad (2)$$

Following our point source paradigm, we define the PSF to be the power spectrum of plane waves in the pupil plane

$$\Phi(\boldsymbol{\kappa}) = |\Psi(\boldsymbol{\kappa})|^2. \quad (3)$$

Our image plane is a scaled version of the angle $\boldsymbol{\theta} = \boldsymbol{\kappa}/k$, which we subsume into the notation for the sake of clarity.

The OTF[1] arises when considering the relationship between the object plane intensity and image plane intensity in a system where the PSF translates linearly with the source position in the object plane, but does not change its shape in any way. With incoherent light in the object plane, overlapping PSFs do not interfere, resulting in an image intensity that is the convolution of the PSF with the object plane intensity. This convolution is mathematically equivalent to multiplication by a complex 2-D transfer function (the OTF) in the spatial frequency domain. We will define our OTF as the inverse Fourier transform of the PSF

$$\mathcal{O}(\boldsymbol{\xi}) = \frac{1}{(2\pi)^2}\int e^{-i\boldsymbol{\kappa}\cdot\boldsymbol{\xi}}\Phi(\kappa)\mathrm{d}^2\kappa. \quad (4)$$

Following the sensible definition of the OTF as worded here would suggest that it be defined as the forward Fourier transform of the PSF rather than the inverse. However, since the OTF has a simple Hermitian symmetry and will be measured in units that involve photon fluxes and sensor counts, we will define the OTF using the inverse Fourier transform to avoid having to carry factors of $(2\pi)^2$ in all the subsequent OTF and dOTF formulae. This would alter the incoherent imaging formula in a trivial way, but we will never need to write it in this article.

Note that since the OTF is the Fourier transform of a real function, it has Hermitian symmetry

$$\mathcal{O}(\boldsymbol{\xi}) = \mathcal{O}^*(-\boldsymbol{\xi}). \quad (5)$$

Combining our definition of the OTF [Eq. (4)] with Fourier optics [Eqs. (2) and (3)] allows us to write

$$\mathcal{O}(\boldsymbol{\xi}) = \int \psi(\mathbf{x}' + \boldsymbol{\xi}/2)\psi^*(\mathbf{x}' - \boldsymbol{\xi}/2)\mathrm{d}^2x'. \quad (6)$$

This formula expresses the OTF as the spatial average of the MCF with baseline $\boldsymbol{\xi}$ over the pupil plane. This integral is concisely written as the autoconvolution of the pupil field

$$\mathcal{O}(\boldsymbol{\xi}) = \psi * \psi^*. \quad (7)$$

### 2.1 The dOTF

The motivation for this technique is the generic concept that differentiating a quadratic function leads to a linear one.[8] The PSF, or equivalently the OTF, is a quadratic functional of the pupil field and can be mostly reduced to a linear functional by performing a non-analytic functional derivative with respect to the pupil mask.[7,8] However, this abstract mathematical approach tends to obscure the simplicity of the derivation when used with Fourier optics and shift-invariant PSFs. Therefore, we shall restrict the present discussion to a direct algebraic treatment.

We start by taking two PSF images, one with the pupil in its default configuration and another with a modification introduced within a small region of the pupil mask. Each PSF is inverse Fourier transformed to find the OTF and subtracted. We call this quantity the dOTF. When we alter the pupil mask, we make the change

$$\Pi(\mathbf{x}) \to \Pi(\mathbf{x}) + \delta\Pi(\mathbf{x}), \quad (8)$$

which causes the pupil field to change

$$\psi(\mathbf{x}) \to \psi(\mathbf{x}) + \delta\psi(\mathbf{x}). \quad (9)$$

Changing the pupil field causes the OTF to change by

$$\delta\mathcal{O}(\boldsymbol{\xi}) = \mathcal{O}_{\Pi+\delta\Pi}(\boldsymbol{\xi}) - \mathcal{O}_{\Pi}(\boldsymbol{\xi})$$
$$= (\psi + \delta\psi) * (\psi + \delta\psi)^* - \psi * \psi^* \quad (10)$$
$$= \psi * \delta\psi^* + \delta\psi * \psi^* + \delta\psi * \delta\psi^*. \quad (11)$$

Writing this out as integrals we find

$$\delta\mathcal{O}(\boldsymbol{\xi}) = \int \psi(\mathbf{x}' + \boldsymbol{\xi}/2)\delta\psi^*(\mathbf{x}' - \boldsymbol{\xi}/2)\mathrm{d}^2x'$$
$$+ \int \delta\psi(\mathbf{x}' + \boldsymbol{\xi}/2)\psi^*(\mathbf{x}' - \boldsymbol{\xi}/2)\mathrm{d}^2x'$$
$$+ \int \delta\psi(\mathbf{x}' + \boldsymbol{\xi}/2)\delta\psi^*(\mathbf{x}' - \boldsymbol{\xi}/2)\mathrm{d}^2x'. \quad (12)$$

If we call the dOTF terms that are linear in $\delta\psi$

$$\delta\mathcal{O}_\pm(\boldsymbol{\xi}) = \int \psi(\mathbf{x}' \pm \boldsymbol{\xi}/2)\delta\psi^*(\mathbf{x}' \mp \boldsymbol{\xi}/2)\mathrm{d}^2x', \quad (13)$$

the OTF's Hermitian symmetry requires

$$\delta\mathcal{O}_+(\boldsymbol{\xi}) = \delta\mathcal{O}_-^*(-\boldsymbol{\xi}). \quad (14)$$

Calling the quadratic $\delta\psi$ term

$$\delta\mathcal{O}_{\delta\delta}(\boldsymbol{\xi}) = \int \delta\psi(\mathbf{x}' + \boldsymbol{\xi}/2)\delta\psi^*(\mathbf{x}' - \boldsymbol{\xi}/2)\mathrm{d}^2x' \quad (15)$$

allows us to concisely write the dOTF as

$$\delta\mathcal{O}(\boldsymbol{\xi}) = \delta\mathcal{O}_+(\boldsymbol{\xi}) + \delta\mathcal{O}_+^*(-\boldsymbol{\xi}) + \delta\mathcal{O}_{\delta\delta}(\boldsymbol{\xi}). \quad (16)$$

### 2.2 Compact Pupil Modifications and dOTF Regions

If the region over which the pupil is modified has sufficiently compact support, i.e., $|\delta\Pi(\mathbf{x})| > 0$ for $|\mathbf{x} - \mathbf{x}_0| \leq \rho$ and $\rho$ is small compared to any spatial structures in the local field, then $\delta\psi$ "samples" the local field in Eq. (13), giving the approximate result

$$\delta\mathcal{O}_+(\boldsymbol{\xi}) \approx \psi(\mathbf{x}_0 + \boldsymbol{\xi})\mu^*, \quad (17)$$

where

$$\mu = \int \delta\psi(\mathbf{x}')d^2x'. \quad (18)$$

This gives us the approximate dOTF in the case of a compact pupil modification as

$$\delta\mathcal{O}(\boldsymbol{\xi}) \approx \mu^*\psi(\mathbf{x}_0 + \boldsymbol{\xi}) + \mu\psi^*(\mathbf{x}_0 - \boldsymbol{\xi}) + \delta\mathcal{O}_{\delta\delta}(\boldsymbol{\xi}) \quad (19)$$

which is useful when $\rho$ is small compared with the inverse spatial bandwidth of $\Phi(\boldsymbol{\kappa})$. That is, if the width of the angular spectrum of the pupil field is $\theta_0$, then we can use the approximate formula if $\rho \ll \lambda/\theta_0$. Similarly, if the PSF is detected out to a radius of $\theta_0 \sim n\lambda/D$, where $D$ is the pupil diameter, then we need $\rho \ll D/n$ for Eq. (19) to be strictly accurate. Both in practice and conceptually, the approximate form can be useful well beyond that limit.

Each of the three terms in Eqs. (16) or (19) occupy specific regions of the $\boldsymbol{\xi}$ plane, as illustrated in Fig. 2. The linear terms are offset such that the pupil modification is at $\boldsymbol{\xi} = 0$, with the two conjugate terms reflected about the modification. If the modification is located at an outermost point of the pupil, the two linear terms will have minimal overlap. While the individual terms have no particular symmetry due to the arbitrary nature of the pupil field, anti-Hermitian symmetry (anti-symmetric real part and symmetric imaginary part) about the pupil modification will cancel between the two terms in the overlap region. The quadratic term is a mini-OTF of the changed pupil field, and is always centered on $\boldsymbol{\xi} = 0$ and contained within the pupil overlap region.

If we only consider the unobscured pupil region described by $\delta\mathcal{O}_+$, we can estimate the phase of the pupil field as

$$\hat{\phi}(\mathbf{x}) = \arg\{\delta\mathcal{O}(\mathbf{x} - \mathbf{x}_0)\} \\ + \arg\{\mu\} - \arg\{\Pi(\mathbf{x})\} \quad (20)$$

and the amplitude $\hat{\alpha}(\mathbf{x})$ as

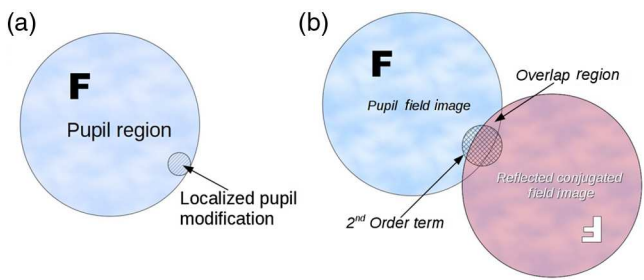

**Fig. 2** Regions in the dOTF. (a) The pupil and the small region where the pupil mask is changed in some way. (b) The dOTF has three terms: the pupil field, the reflected and conjugated pupil field, and the localized contribution of the second-order term. The field in the overlap and quadratic term are not useful for measuring the wavefront, and the reflected term gives redundant information. However, by placing the pupil modification near the edge, we can directly estimate the field over most of the pupil. A different placement of the pupil modification would yield the same pupil field, but the overlap region would be elsewhere in the pupil.

$$\hat{\alpha}(\mathbf{x})|\mu\Pi(\mathbf{x})| = |\delta\mathcal{O}(\mathbf{x} - \mathbf{x}_0)|. \quad (21)$$

Extended pupil modifications can bias both of these estimates through the convolution in Eq. (13).

### 2.3 A Numerical Example

As an illustrating example, we will use the monolithic mirror telescope (MMT, http://www.mmto.org) pupil with $D = 6.5$ m using the $f/15$ 640 mm secondary obstruction. The choice of wavelength itself is not particularly important to this calculation other than affecting the grid spacings and the amount of phase wrapping, so we arbitrarily choose a mid-infrared wavelength of 3.8 $\mu$m (L-band) where such an experiment might feasibly be performed. The incoming starlight wavefront is assumed to have passed through Kolmogorov turbulence resulting in a Fried length $r_0 = 15$ cm at $\lambda = 550$ nm, corresponding to $r_0 = 1.5$ m at $\lambda = 3.8$ $\mu$m. This gives $D/r_0 = 4.3$ which is easily calculated with reasonable sampling in both the pupil and image spaces. We chose a spatial sampling of 4 cm and the PSF was computed to a radius of 1 arcsec with a generously small pixel size of 0.01 arcsec (i.e., a plate scale of $\sim 1/6$ Nyquist). The PSF actually only needs to be sampled at or better than Nyquist, so that the most distant point in the pupil from the pupil modification does not overlap with the aliased version. (A sub-Nyquist-sampled PSF can still fully map the pupil field by choosing at least one more pupil modification position and taking at least one more image. The obscured regions would rearrange, allowing a complete pupil field to be pieced together from the two dOTFs.) For our calculation, the pupil was modified by placing a $4 \times 12$ cm$^2$ mask at the pupil's edge. This choice of aspect ratio would be sub-optimal for a real observation and was only selected here to make the effects of the convolution with $\delta\psi$ more obvious.

The pupil fields were created from $\psi(\mathbf{x}) = \exp\{i\phi(\mathbf{x})\}\Pi(\mathbf{x})$ and the PSFs computed using magnitude-squared, zero-padded FFTs. With such a small grid spacing (the maximum angle is $\pm 9.8$ arcsec which is $162.5\lambda/D$), we can expect numerical artifacts to be minimal. Note that the total flux passing through the two pupils is not expected to be the same because their areas differ. In a real measurement, we would have to be careful about relative normalization since the incoming photon flux can vary. We next inverse FFT each of the PSFs and take the difference between them to find the dOTF. Figure 3 shows the baseline (a) and modified (b) pupils. The random Kolmogorov phase screen used to create the pupil field is shown in Fig. 3(c). The magnitude of the dOTF is shown in Fig. 3(d). Notice that while the compact modification limit of the dOTF [Eq. (21) with Eq. (17)] suggests that the amplitude should be constant in the non-overlapping pupil regions, while instead it appears mottled due to the convolution with $\delta\psi$. In this example the effect is not extreme, but it is noticeable; it just leaves a schlieren-like residual of the phase variation. The most obvious effect of the convolution is to smear the field in the direction of the obstruction's axial asymmetry, clearly seen in Fig. 3(d) when looking at the edges of the pupil and spider vanes. Finally, the unwrapped phase of the dOTF is shown in Fig. 3(f). The phase unwrapping algorithm was the unwt "unweighted least-squares method of phase unwrapping by means of direct transforms" from

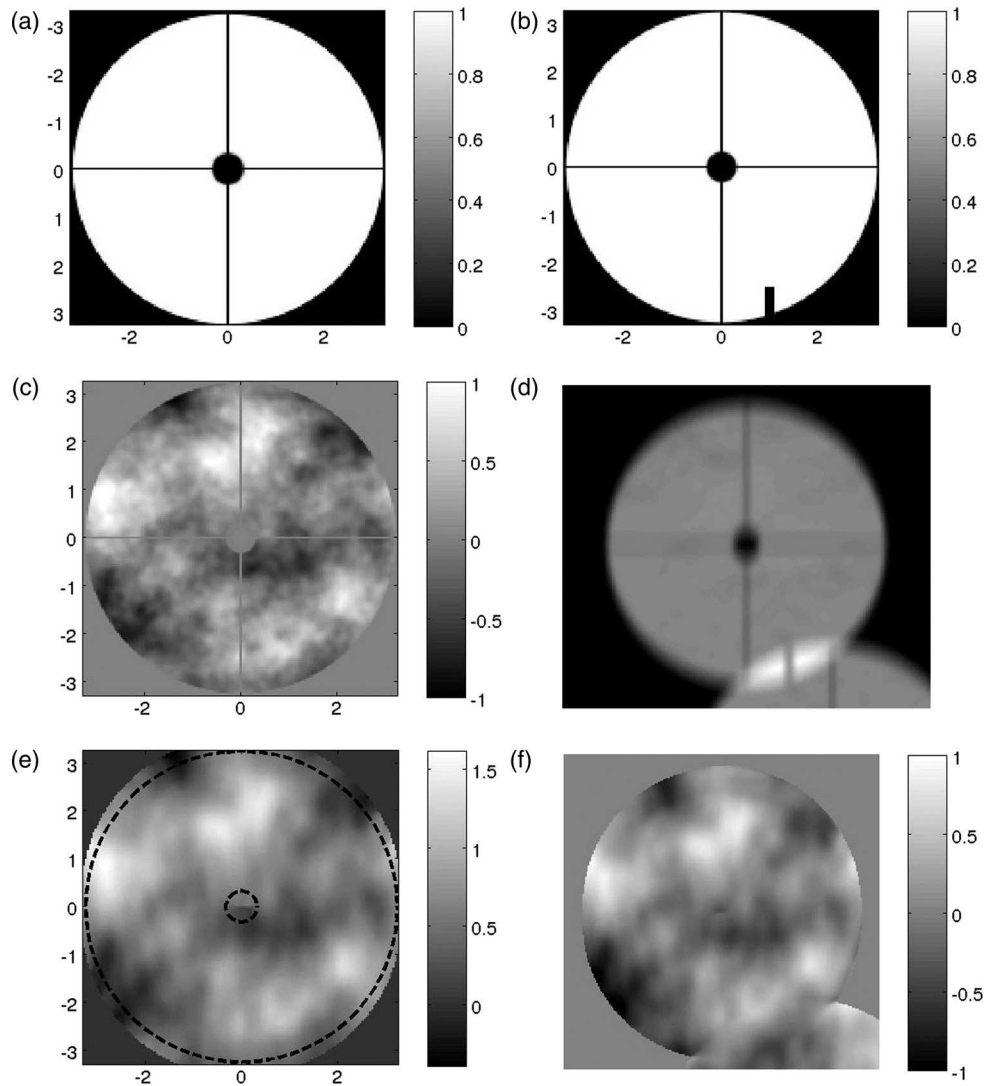

**Fig. 3** A numerical example of the dOTF. The monolithic mirror telescope (MMT) pupil is shown in (a). The pupil modification is shown in (b). It is an elongated rectangle to more clearly show the effects of the $\delta\psi$ convolution. A Kolmogorov atmospheric phase pattern is shown in (c). The calculation proceeds by computing the PSF for the two pupil masks, Fourier transforming, and computing the difference. The magnitude of the dOTF is shown in (d) and the unwrapped phase is shown in (f). The phase of the true pupil field convolved with the pupil difference [(b)–(a)] is shown in (e). The actual pupil extent is indicated by the dashed circle.

Ghiglia and Pritt.[10] This would have been the actual phase of the pupil field had the dOTF been measured with a sufficiently compact pupil modification. Instead, it is smeared in a way that is similar to the amplitude, but is more complicated due to the non-linearities involved in convolving against the complex exponential. A leading-order analysis of the effect of the convolution on the phase estimation formula, Eq. (20), suggests that the estimated phase $\hat{\phi}(\mathbf{x})$ for a weak turbulence-like random phase input is mostly unbiased, but loses higher spatial frequencies similar to convolving the actual phase with $\delta\psi^*$. The recovered pupil phase may be biased within a distance $\rho$ of the pupil boundary, especially if $\delta\psi$ contains significant phase structure. Figure 3(e) is the argument of the actual pupil field $\Pi(\mathbf{x})\exp[i\phi(\mathbf{x})]$ convolved with $\delta\Pi^*$ for comparison. The dashed circles indicate the nominal pupil boundaries. Comparison of "truth" with the dOTF estimate is quite good, but a detailed accuracy analysis is beyond the scope of this introductory example. If greater precision is required, deconvolution may be a viable option, depending on how well the pupil change is known and on the impact of any unknown spatial variations of the incident field across the pupil modification.

### 2.4 The Effect of Optical Bandwidth

The dOTF measurement accuracy increases with more detected photons, so for non-laser applications we will find it advantageous to use larger optical bandwidths. The effect of bandwidth is computed by incoherently summing narrowband PSFs. The resulting reference and modified PSFs are radially blurred by the usual $\theta = (\lambda/\lambda_0)\theta_0$ scaling with longer wavelengths appearing farther from the optical axis. When we compute the OTFs by Fourier transforming the broadband PSFs, each wavelength scales inversely with the effect that all wavelengths contribute nearly the same function, but with progressively smaller radial scale factors for longer wavelengths. The result is a radial blurring of the OTF similar to the PSF but in reverse. Upon differencing, the radial smearing in the dOTF will be centered on the pupil modification and will have a radial extent of $\delta\boldsymbol{\xi} = (\delta\lambda/\lambda_0)\|\boldsymbol{\xi}\|$. Thus, a 5% bandwidth will lead to a radial blurring at the most distant point in the pupil that is 5% of the pupil diameter with points closer to the modification being affected proportionally less. This can either be used to set a maximum bandwidth or the radial blurring can be compensated by using some scaled form of radial deconvolution. (A proportional radial deconvolution can be performed by using a log-polar mapping: the complex dOTF image in $x - y$ is mapped to a complex function in the stretched polar coordinates of $\theta - \log r$. In this space, the proportional radial blurring is one-dimensional and translation independent, making deconvolution relatively straightforward using conventional techniques. Once processed, the deconvolved log-polar dOTF would be transformed back into $x - y$ coordinates.) An interesting consequence of the proportional radial blurring is that we can combine several dOTF estimates with pupil modification points distributed around the edge of the pupil to create a composite dOTF estimate that is less affected by bandwidth. For example, by combining 3 or 4 dOTFs from pupil modification at various points around the pupil, and keeping each result only out to, say, half the diameter of the pupil, the combined dOTF will only have half the maximum blurring appearing in the center of the merged pupil field result.

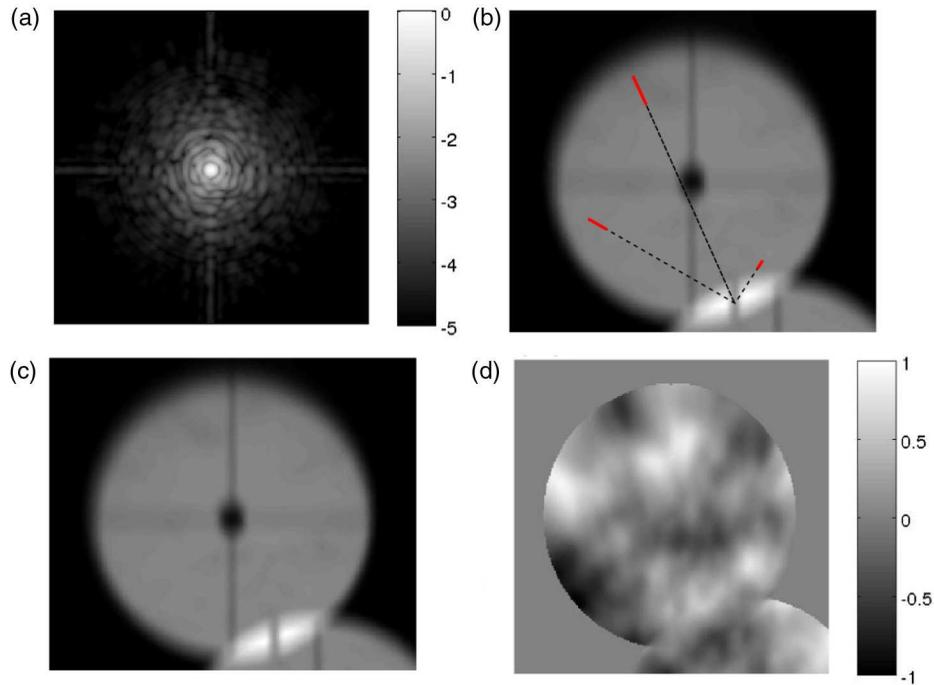

**Fig. 4** Effect of a 10% bandwidth on the MMT dOTF example in Fig. 3. (a) The MMT PSF computed from the assumed pupil phase aberrations and a 10% optical bandwidth. The resulting PSF is radially smeared by 10% in radius. (b) Radial smearing in the PSF transforms to radial smearing in the dOTF with the color sequence reversed, the radial smearing being centered on the pupil modification. A 10% bandwidth has a negligible effect near the point of modification, but causes a larger blurring effect at more distant points in the pupil. (c) The amplitude of the dOTF shows the effects of both the rectangular pupil mask and the radial blurring due to bandwidth. (d) The dOTF phase [compare with Fig. 3(c) and 3(f)].

The radial bandwidth smearing is shown in Fig. 4 by adding a 10% bandwidth to the case modeled in Fig. 3. The simulated phase pattern used above is assumed to arise from a non-dispersive atmospheric turbulence or a geometric wavefront distortion, such that the phase pattern scales with wavelength as $\phi(\mathbf{x}, \lambda) = \phi(\mathbf{x}, \lambda_0)\lambda_0/\lambda$. The individual narrowband PSFs were computed every 1% in $\lambda/\lambda_0$ with the appropriately scaled phase patterns and then incoherently summed with a constant (white) spectral energy distribution. As a result, the PSFs are radially smeared by 10% in radius from the center of the pupil modification. Figure 4(b) shows $|\delta\mathcal{O}|$ with various points called out, red segments indicating a 10% radial averaging window. Figure 4(c) shows $|\delta\mathcal{O}|$ and Fig. 4(d) shows $\arg\{\delta\mathcal{O}\}$. Notice that near the point where the pupil modification was introduced, the effect of bandwidth is relatively slight and the resolution is dominated by convolution with $\delta\psi^*$. Farther from the modification, the effect of bandwidth becomes more pronounced, but the resolution is not bandwidth limited until the bandwidth blurring exceeds the extent of the pupil modification: $\|\boldsymbol{\xi}\|\delta\lambda/\lambda_0 \gtrsim \rho$.

### 2.5 Possible Implementations and Practical Considerations

In the above example, the pupil was changed using a transmission blockage placed over a small peripheral area. If this option is possible and satisfies reasonable constraints of time and PSF stability, then it is a very simple and effective method for making a dOTF measurement. It is also possible to use phase to modify the pupil. For example, if the system includes a DM or a spatial light modulator (SLM), then a simple displacement of an actuator near the edge of the pupil will allow nearly all the pupil field to be measured. Note, however, that the phase modifications can lead to more extreme convolution effects. Instead of the dOTF giving the pupil field convolved with a relatively simple, nominally real function, the phase-induced dOTF consists of the pupil field convolved with $\{\exp[-i\varphi(\mathbf{x})] - 1\}\alpha(\mathbf{x})\exp[-i\phi(\mathbf{x})]$, where $\varphi(\mathbf{x})$ is the phase change introduced by the DM between images. If the maximum phase shift is larger than $\pi$, the convolution kernel will have an oscillatory character that will cause more extreme edge detection properties. Also, since the mean value $\mu$ may be reduced by the oscillations, the overall dOTF may be reduced in magnitude, resulting in a lower sensitivity (Fig. 5).

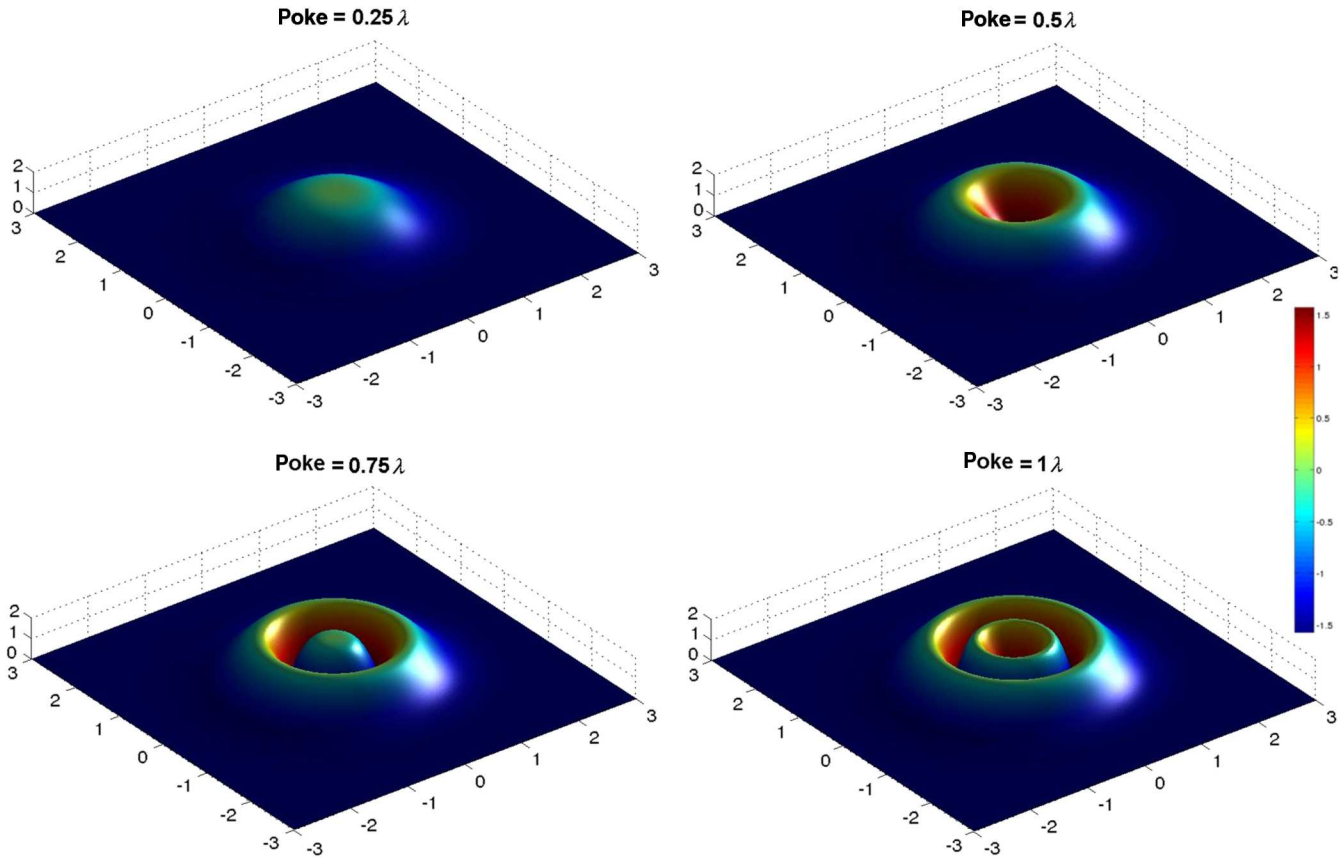

**Fig. 5** The change in the pupil field $\delta\psi$ for a Gaussian DM poke. Each of the surfaces represent the complex field amplitude as height with phase as color. A modest poke is an effective probe while deeper pokes create oscillatory fields that cause the dOTF to be a more significantly spatial filtered version of the pupil field. The poke displacement is half the down-and-back change in the wavefront. A $\lambda/4$ poke (upper left) is the largest displacement possible before field wrapping begins to occur. By the time the poke depth is a full wavelength (lower right), $\delta\psi$ has two rings and the center value is zero. A poke of $<\lambda/4$ is best.

This effect can be seen when considering how far a single actuator should be displaced in order to achieve the best dOTF signal. We start with Eq. (13) and an incident plane wave, $\psi_0 = 1$, including an actuator displacement of $\ell$ at $\mathbf{x}_0$ with a generic influence function $z(\mathbf{x}) = \ell g(\mathbf{x} - \mathbf{x}_0)$, then $\delta\Pi(\mathbf{x}) = \exp\{2ik\ell g(\mathbf{x} - \mathbf{x}_0)\} - 1$. (The factor of 2 is due to the fact that the DM is a mirror and the down-and-back propagation distance is twice the surface displacement.) For simplicity, we will use a top hat influence function of radius $\rho$. The resulting dOTF amplitude is $|\delta\mathcal{O}^+| \propto \pi\rho^2 \sqrt{1 - \cos 2k\ell}$, which has its first maximum when $\ell = \lambda/4$ and scales as the area of the actuator influence. A more realistic influence function will scale similarly, but oscillate around its edge when $\ell > \lambda/4$ with no increase in signal (Fig. 5). Therefore, for ease of interpretation with the best spatial resolution and maximum signal, a DM actuator displacement of $\lambda/4$ or less is preferred.

The leading practical considerations for measuring the dOTF include PSF saturation, vibration, and signal-to-noise. PSF saturation is an issue because it causes a broad error term to appear in each OTF, resulting in an error that becomes more significant farther from the point of pupil modification where the MTF drops closer to the noise floor. A high-quality dOTF measurement requires a sufficient number of photons in each of the PSF exposures for the dOTF signal to dominate other noise contributions. In a controlled laboratory setting, photons are generally available in abundance, allowing for very short exposures. However, typical image sensor well depths are such that the PSF peak rapidly saturates and possibly overflows. Since saturation will spoil the Fourier transforms, the flux must be attenuated or exposure time reduced to keep the brightest parts of the PSF on-scale and linear. If the frame rate of the camera is limited, this can seriously slow the process of collecting the required total number of photons. Using pixels that are much smaller than the required Nyquist sampling can help, but may not be possible and increases read noise. A very simple technique for decreasing the total required exposure time is to introduce some defocus, possibly a significant amount.[11] This reduces the Strehl ratio, dimming the brightest parts of the PSF and spreading the light over a larger area of the sensor: allowing many more photons to be captured in a single unsaturated exposure. The effect of defocus on the dOTF is to add a quadratic phase profile across each of the pupil field images. This profile is easily estimated and removed, leaving the same dOTF that would have resulted from a properly focused PSF. Defocus can increase the number of photons that can be detected in a single unsaturated exposure by 2 or 3 decades, significantly increasing the achievable signal. Note that extreme defocus will also induce a significant phase gradient across the pupil modification, causing amplitude and phase artifacts to appear in the dOTF. When the diffraction pattern of the defocus-tilted pupil modification and the overall pupil field's diffraction pattern no longer overlap, the dOTF will lose signal. This sets a practical limit to the use of defocus unless efforts are made to compensate for the defocus tilt with a tilt across the pupil modification.

Vibration causes the OTF estimates to include an extra mutual coherence factor that is unlikely to be common between the reference and modified images, leaving un-subtracted power outside the nominal pupil regions in the dOTF. Noise, regardless of its source, will cause random errors in both the dOTF phase and amplitude. If the SNR is unacceptable, it is possible to bin "pixels" in the complex dOTF before taking the argument or the absolute value, thereby increasing the accuracy of the field estimate at the expense of spatial resolution. The best precision will be achieved when the binned pixel size is matched to the blurring caused by the $\delta\Pi$ convolution.

The derivation of the dOTF assumes that the change in the pupil field is spatially localized. If there is an additional change that is broadly distributed over the pupil, e.g., an evolving turbulent phase pattern, an actively changing AO DM, or even a small vibration affecting only tip-tilt, the OTFs will not subtract properly, leaving an OTF residual that can overwhelm the dOTF signal. Attempts to measure the dOTF's full and modified aperture PSFs simultaneously using some sort of a beam splitter arrangement can easily incur fixed NCP aberrations that are difficult or impractical to remove from the contaminated dOTF, even with calibration. Acquiring the PSFs sequentially over the same optical path eliminates fixed NCP errors, but can introduce temporal NCP errors from vibration or OPL variations from air turbulence. While vibration can be dealt with by taking a series of short exposures and applying a tip-tilt correction to the individual frames, more general phase differences are more troublesome. A full derivation of the effect of temporal phase fluctuations during the individual PSF exposures is beyond the scope of this article, but the effects are easily summarized. The most obvious effect is an extended residual that does not properly subtract beyond the limits of the dOTF pupil image regions. In long PSF exposures, the individual OTFs incur an additional "seeing factor" of $\exp\{-D_\varphi(\xi)/2\}$ where $D_\varphi(\xi) = \langle[\varphi(\mathbf{x}+\xi/2) - \varphi(\mathbf{x}-\xi/2)]^2\rangle$ is the varying phase structure function during the exposures. This causes a loss of dOTF signal farther from the pupil modification, leading to noisier phase measurements and reduced amplitude measurements. If the rms phase fluctuations across the pupil diameter are larger than 1 rad, this effect can be significant. Shorter exposures that do not fully average over the ensemble of phase fluctuations lead to errors that act more like static NCP errors. An effective method for mitigating temporal NCP error is to approximate simultaneous full and modified pupil exposures by multiplexing. In this approach, interleaved short exposures are taken with alternating full and modified pupils, which are then summed and differenced appropriately. The remaining temporal NCP effects are determined by the extraneous pupil field changes that occur during a single short exposure pair as opposed to the amount incurred between two fully non-overlapping exposures.

## 3 Experiments

The dOTF theory can be demonstrated using an adequately sampled imaging system and any pupil modification that is available. The following experiments are presented as qualitative proofs of concept. We consider two cases: a local pupil transmission change introduced using a small blockage and a phase change introduced by moving a single actuator in an MEMS DM.

### 3.1 A Simple Pupil Blockage

We first consider a small pupil blockage. A HeNe laser was focused with a microscope objective onto a (nearly extraneous) 50 $\mu$m pinhole (Fig. 6). The diverging beam was collimated and passed through an iris, set to a diameter of ~5 mm. The light was reflected off two plane mirrors to fit onto the breadboard and finally brought to an image using a 300-mm focal length lens. The laser light was imaged using a 10-bit Firefly FFMV-03M2M IEEE1394 (Firewire) CMOS camera from Point Grey, Inc, Richmond, BC, Canada. The camera has a 640 × 480 array of 6 $\mu$m square pixels and was read out with a gamma of unity. The plate scale was measured (using the diameter of the MTF relative to the Nyquist frequency) to be $0.19\lambda/D$ per pixel or 2.7 times finer than the Nyquist sampling. The pupil blockage was introduced using a short length of 24 gauge wire attached to a tip-tilt stage mounted on a translation stage. The wire was adjusted to be close to the iris and adjusted by hand during the experiment (Fig. 7).

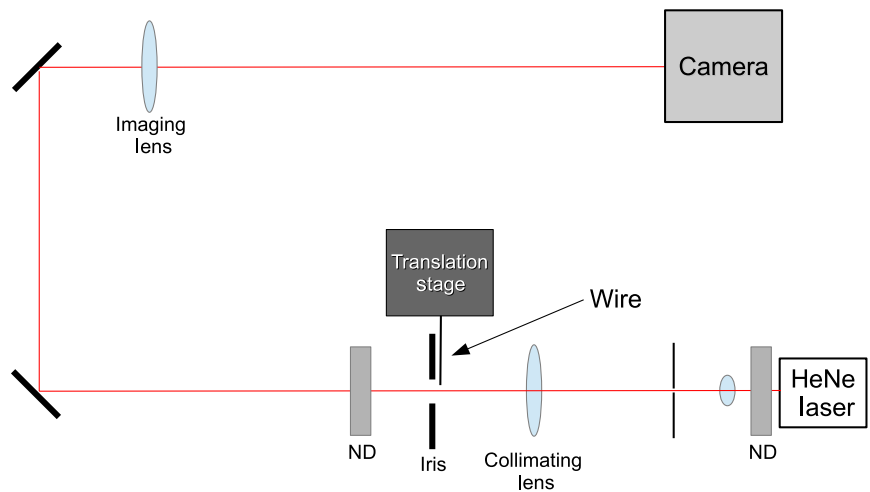

**Fig. 6** Schematic for the simple pupil blocker experiment.

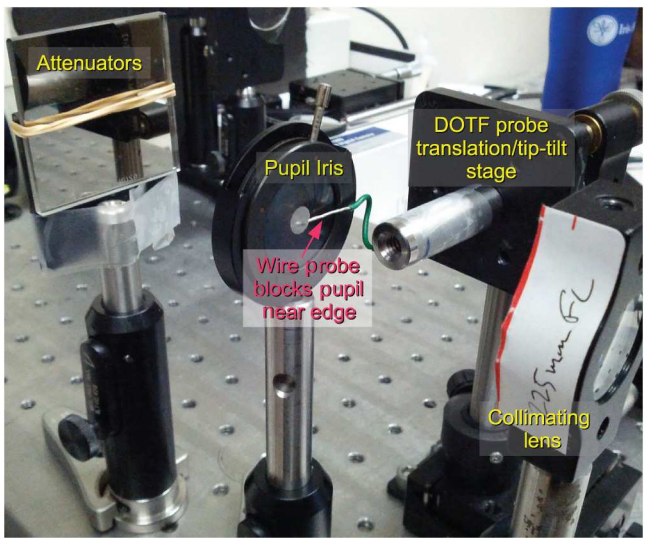

**Fig. 7** Pupil block detail. A simple piece of wire was attached to a translation/tip-tilt stage and used to block a small area of the pupil near the edge.

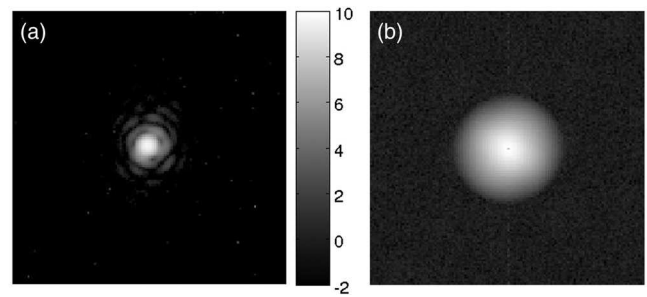

**Fig. 8** An example PSF from the pupil blocking case (a) and the corresponding MTF (b). The PSF is shown in base-2 log and the scale is in bits. The camera is 10 bits and the image is based on the average of 25 images. The MTF is computed from magnitude of the OTF, and the image is the MTF to the 1/4 power to greatly reduce contrast. The noise down the centerline is due to a row noise artifact in the camera.

The image was adjusted for best focus by maximizing the peak intensity. In order to avoid saturating the PSF, multiple ND filters were placed in the beam and the exposure time increased until the peak fell between 70% and 80% of the maximum value (1024). The final exposure time was 356 $\mu$s or approximately 1% of the observation time when continuously read out at 30 fps. Data was collected at full video frame rate, 25 frames at a time, which were then averaged into a single image and kept for further processing. A set of dark images was also acquired and the average was subtracted from the later PSF frames. This step may appear to be unnecessary since the dOTF is linear in intensity, and the dark subtraction step explicitly cancels out. However, if the laser power had been varying, requiring a scaling of one OTF relative to the other before subtraction for the dOTF, not having first performed a dark subtraction would introduce noise. The baseline PSF was first imaged and Fourier transformed to find the baseline OTF. Since the wavefront was fairly flat, the OTF was close to real over the entire area $|\boldsymbol{\xi}| < D$ [Fig. 8(b)]. The breadboard system was constructed on a regular desktop without any attempt at vibration control, and the series of short exposures provided the opportunity for vibration correction on a per-frame basis. The PSF was clipped out of each full CCD image frame and placed at the center (65, 65) of a $128 \times 128$ sub-image. The peak was then circularly shifted to the (1, 1) pixel before performing a 2-D FFT and circular-shifting back. An example PSF and MTF (magnitude of the OTF $= |\mathcal{O}|$) are shown in Fig. 8. In addition to the basic processing, the PTF $= -\arg\{\mathcal{O}\}$ was computed from the OTF and fit to a plane as a function of $\boldsymbol{\xi}$. This fit then allowed the application of a vernier tip-tilt correction to the PSF in the OTF space, accurate to a fraction of a pixel. Although this processing step was included, vibration was not an issue and could have been neglected in this case. No effort was made to correct for flux variations since the laser flux was quite stable.

Using averaged blocks of 25 vibration-stabilized video frames, the dOTF was computed using $\delta\mathcal{O} = \mathcal{O}_{\mathrm{mod}} - \mathcal{O}_0$ and displayed as shown in Fig. 9. Column (a) in Fig. 9 is a complex image of the dOTF, where color represents phase and brightness represents a gamma-scaled version of the complex amplitude. The second and third columns show the same information called out separately: column (b) is $|\delta\mathcal{O}|$ and (c) is the derived wavefront displacement $z = \arg\{\delta\mathcal{O}\}\lambda/2\pi$. The grayscale is linear and set to run over 300 nm from white to black. The resulting dOTF images were acquired, processed, and displayed at about 1 frame/s, but could easily be increased to full video frame rates using the defocus method to allow a greater number of photons to be detected per frame (Sec. 2.5). Even at 1 frame/s the result was striking. The intensity variation across the pupil appears to be real, the result of a too large point source pinhole and too short of a propagation before collimation. The wavefront itself (i.e., the phase measured in wavelengths) was very stable and moved the dOTF as expected when the placement of the blocking wire was adjusted. Although no attempt was made to dynamically correct the wavefront using only the dOTF (as it would have required continuously taking new reference PSFs, which would have been cumbersome with the simple setup), we did compute the single measurement accuracy from the collection of images (Fig. 10). Using 60 measurements, each consisting of 25 video exposures, we computed the phase using $\phi = \arg\{\delta\mathcal{O}\}$. The standard deviation was then computed per-pixel [Fig. 10(b)] and found to range from about 0.04 to 0.07 rad rms. This corresponds to a 1-sigma error of about 6 nm per each second of measuring. From this we can extrapolate how long it would take to measure the wavefront to 1 nm rms, assuming we took equal numbers more of reference and mod images, at the same inefficient frame rate and exposure duty cycle. Assuming unbiased normal statistics, we would need $2 \times 6^2 = 72$ images, each of which are 25 video frames. With our simple setup, it would take 1 min to reach 1 nm rms. Again, this time could be significantly reduced with some defocus and better management of the video stream. Even so, the performance was excellent and far easier to set up and operate than an interferometer capable of making the same accuracy measurement.

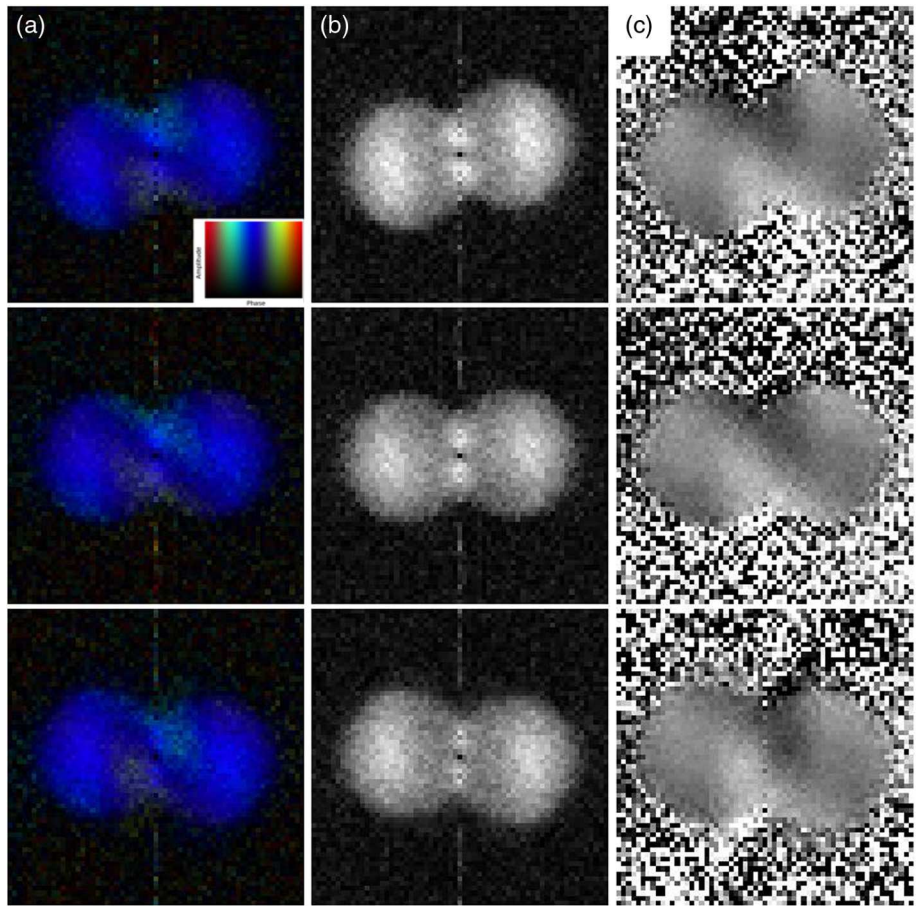

**Fig. 9** Simple pupil blocker dOTF results. The three rows contain examples of the dOTF with the probe wire placed in slightly different positions along the edge of the pupil and with different insertion depths. The third case has the smallest insertion and therefore has the smallest pupil conjugate overlap. The first column (a) contains a complex image of the dOTF where brightness indicates amplitude and color indicates phase (see inset key). The middle column (b) shows the magnitude of the dOTF, nearly proportional to the pupil field amplitude in this case. Finally, the third column (c) shows the pupil wavefront, with the grayscale running from $-150$ nm to $+150$ nm. By computing $\exp\{-\langle(\varphi - \langle\varphi\rangle)^2\rangle_{\delta\mathcal{O}\neq 0}\}$ the Strehl ratio was estimated to be 87%.

### 3.2 dOTF with an MEMS DM

The dOTF pupil modification can also be introduced using phase. For example, a localized phase modification can be introduced using a DM by displacing a single actuator. This can be very convenient since in an AO system there are actuators at many points near the edge of the pupil. Displacing even a single actuator near the edge of the pupil can measure nearly the entire pupil field.

In our pupil-blocking tests we did not introduce wavefront test patterns, but simply measured the residual aberrations present in the optical system. The availability of a DM makes it simple to apply phase test patterns, as well as

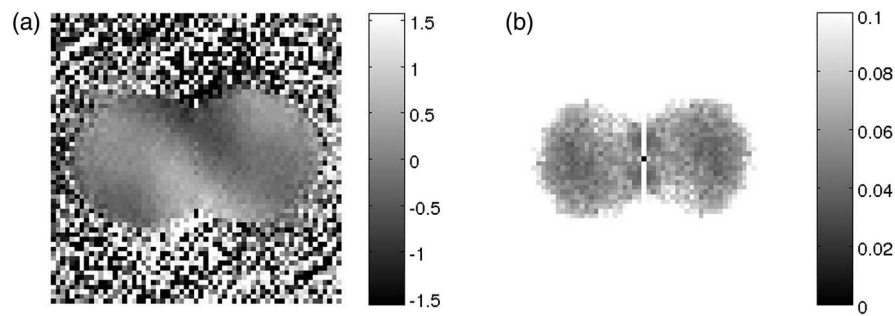

**Fig. 10** dOTF noise and stability estimate. These statistics were derived from 60 measurements of the dOTF, each measured from the PSF mean from 25 video frames. Image (a) is the mean pupil wavefront phase, with the grayscale shown from $-\pi/2$ to $\pi/2$. No phase unwrapping was performed because the wavefront was so nearly flat. The per-frame wavefront phase noise (b) can be estimated by computing the rms phase. This corresponds to a 1-sigma error of about 6 nm rms from 25-coadded frames. With our inexpensive setup, it would take 1 min to reach 1 nm rms.

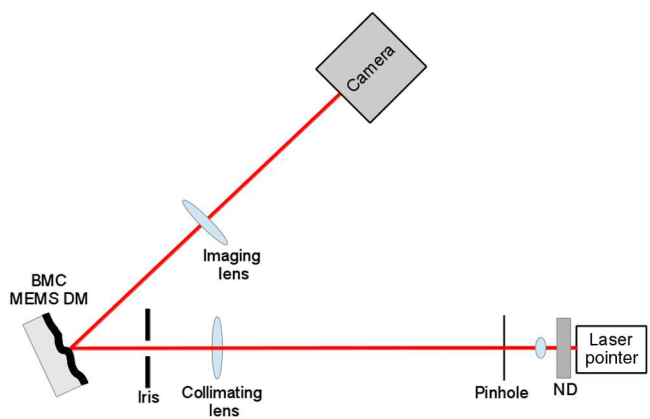

**Fig. 11** MEMS DM "poke" dOTF experiments system diagram.

enabling the pupil modification for a dOTF measurement. A diagram for these experiments is shown in Fig. 11. Light from a red ($\lambda = 635$ nm) diode laser was passed through a microscope objective and a 10 $\mu$m pinhole to create a spatially filtered point source. This light was collimated using a 50 mm focal length lens and passed through a pupil iris where it illuminated a 140-element Boston Micromachines Multi-DM. Finally, a single 150 mm focal length lens was used to focus the light onto the same 10-bit CMOS camera used in the transmission blockage experiment (Sec. 3.1). In this case, the camera pixels were $0.135\lambda/D$ with the pupil iris set to match the 4.4 mm DM. Since the pupil mask obscured some of the corner actuators, an appropriate modification actuator was selected by scanning through the actuators with a deep motion (much more than recommended for the dOTF measurement) and looking for visual changes in the PSF. Once selected, the protocol was to apply a test pattern on the DM, take a baseline image of the PSF, move the probe actuator (ideally by an amount $<\lambda/4$ or approximately 100 to 150 nm), and take a second image. In both the before and after images, the PSF was estimated from the mean of 30 camera frames, each with the shutter timing adjusted to ensure that the PSF was not saturated. Once again, dark images were acquired and subtracted, although without flux correction they drop out of the analysis. The optical breadboard used in this experiment was again placed on a common desktop with no vibration mitigation, and this time the individual PSF images vibrated by several pixels during data collection. The individual exposures were sufficiently short that their PSFs were not blurred by the vibration and could be stabilized using the vernier OTF alignment method described in Sec. 3.1. Once aligned, the video images were averaged and used for further dOTF processing.

Two of the tests are shown in Fig. 12. The first test case, Fig. 12(a), shows the data and dOTF processing for the unpowered DM. The unpowered DM has a relatively smooth shape, but with some optical power. The alignment of the test system was done quickly and therefore does not show a clean Airy pattern. The reference PSF was taken in this configuration, followed by the modified case with one actuator moved by approximately 150 nm. The baseline and modified PSFs are shown in the upper left-hand portion of Fig. 12(a). The PSFs were Fourier transformed to find the corresponding complex OTFs, plotted on the bottom left with intensity representing a gamma-modified map of amplitude and color representing phase (blue is 0, red is $\pi$, and yellow and green are $\pm\pi/2$). The dOTF is the difference between the OTFs and is shown in the center frame of Fig. 12(a). Finally on the right-hand side, the unwrapped phase of the dOTF is shown, converted to DM surface displacement using $z = \phi\lambda/4\pi$. The overall rounded shape of the unpowered DM is clearly visible. Next, in Fig. 12(b), the process is repeated with a single interior actuator poked by 150 nm ($\sim\lambda/4$) for both the baseline and mod images. The resulting dOTF shows the isolated actuator creating a phase difference from the original value (blue or zero phase) to π (red at the peak). This is quantitatively as expected for a poke of 150 nm.

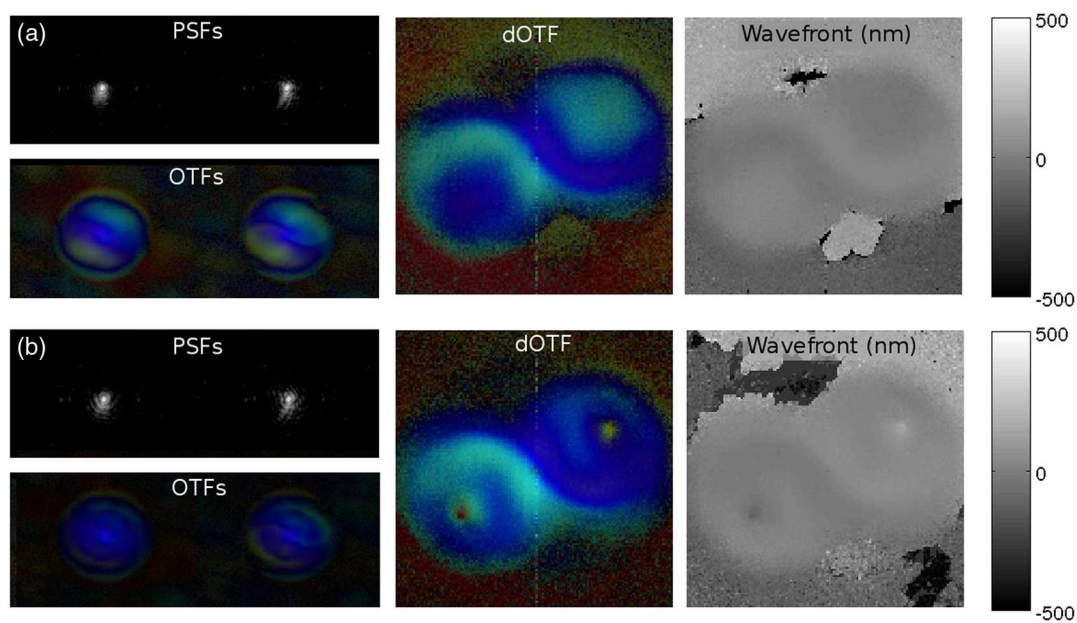

**Fig. 12** "Poke" test with a Boston Micromachines MEMS DM. Case (a) shows the PSFs, OTFs, dOTF, and derived reflecting surface displacement for the unpowered DM. Case (b) shows the corresponding images and fields when a single actuator was poked by ~150 nm ($\sim\lambda/4$).

For the second DM experiment, we set up a problem that would be very difficult for iterative phase retrieval methods with an unknown pupil and unknown details of the pupil transmission mask. The pupil iris was opened to a diameter larger than the 4.4-mm diameter DM and intentionally displaced to one side. The reflectivity of the device varies significantly beyond the mirror's edge with high-angle scattering included. The experiment was intentionally not carefully designed or built, but used available components and an inexpensive laser pointer as a light source to illustrate the robustness of the dOTF technique. The camera pixels sufficiently oversampled the PSF to allow for the larger OTFs from the light reflected beyond the active area of the DM. This time the DM was powered to have a test pattern (the letter "A") with the voltages set to give a maximum pattern relief of about 125 nm (Fig. 13). After poking the edge actuators to find the ones best suited for the pupil modification, actuator (1, 3) was chosen as our probe along the lower edge of the DM on the left leg of the "A" in Fig. 13(b). The baseline PSF is shown in Fig. 14(a) and the modified PSF in Fig. 14(b) with the probe actuator displaced a further $0.4\lambda$ (about 250 nm). A bracketed set of measurements were taken, although we only show one here. The probe actuator displacement was significantly larger than the recommended poke depth but was selected to demonstrate and accentuate the edge detection effects visible in the dOTF magnitude output. The processing was performed as before: the PSFs were dark subtracted, Fourier transformed, and differenced to give the dOTF. Figure 14(c) shows the negative linear grayscale image of $|\delta\mathcal{O}|$, clearly revealing not only the placement of the pupil stop relative to the DM, but the "A" phase pattern. The phase pattern cross-talk from the pupil mod spatial filtering is significantly less prominent in the dOTF amplitude when the probe displacement is smaller (e.g., $<\lambda/4$), but the resulting amplitude signal is so clear that it strongly suggests that the convolution effect be further exploited in future tests and applications. The unwrapped phase (using Ghiglia and Pritt's unweighted phase unwrapping by DCT/FFT algorithm[10]) of the dOTF is shown in Fig. 14(d). The measured pattern depth generally matches the computed applied pattern, although we did not perform a detailed quantitative analysis. With the same flux and light levels shown here, other dOTF test pairs indicated that we were capable of detecting surface height variations of <5 nm in a 1 s measurement, similar to the sensitivity in the pupil blocking case. Using defocus to increase the number of detected photons in a single exposure should enable full video frame rate measurement with the same hardware.

## 4 Possible Applications

The dOTF method is so simple that it can be readily implemented with available components and achieve results comparable to an interferometer. However, since the optical system under test can be measured with dOTF in its operational configuration, including all optical effects and NCP aberrations, it is actually more convenient than an interferometric measurement. The basic dOTF theory describes its use with point sources, Fourier optics, and shift-invariant PSFs, but the technique is amenable to many theoretical extensions and implementation variations. A constellation of point sources with a position-dependent PSF (i.e., non-shift invariant) will all be modified by a single pupil modification. Processing each PSF in the constellation into its own dOTF gives the pupil field as seen along that source's line-of-sight through the optical system, permitting a 3-D tomographic reconstruction of the aberrations and vignetting structures.[12] Because dOTF requires a large number of photons in order to accurately estimate the OTF's change with the pupil modification, it seems unlikely that this technique can be used to replace the conventional WFSs in real-time photon-limited astronomical AO applications. However, for less dynamic measurements, such as the phasing of mirror segments in multisegment telescopes, dOTF may provide a simple and practical alignment technique. It should be possible to build a compact device that measures the phase structure function with a point beacon, enabling simple low-cost turbulence metrology. Any situation amenable to study with an interferometer should also be a potential dOTF application. But, since vibration appears as correctable tip-tilt as opposed to fringe changes, robust measuring systems may be much easier to build. The dOTF technique is likely going to have its most obvious applications in optical shop testing, since it is easy to set up, requires little or no calibration, and gives useful answers with very simple processing. Sophisticated pupil phase retrieval methods, such as those intended for JWST segment phasing, could be supplemented or bootstrapped by a dOTF method where the pupil modification is introduced by blocking a corner of one of the outlying segments with the filter wheel frame or by tilting one of the outlying segments by 100 nm or so. It should be possible to use the dOTF measurements as the first estimate of a phase diversity analysis, speeding convergence. Taken by itself, dOTF can be used to measure NCP aberrations on a longer timescale than the primary AO system and used to drive a slower independent control loop to enhance the Strehl ratio or PSF quality at the science camera.

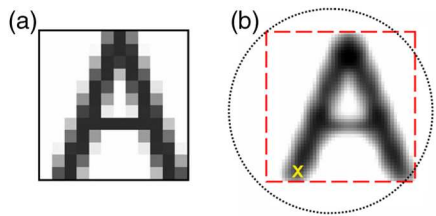

**Fig. 13** The DM test pattern and misaligned pupil illumination. The applied DM test pattern is shown in (a). Once rendered on the DM, the expected DM surface response with a 15% influence and cubic interpolation is shown in (b). The DM active area (dashed) and misaligned pupil illumination area (dotted) are indicated. The yellow "x" indicates the location of the pupil modification, in this case actuator (1, 3), which was displaced roughly 250 nm from the initial pattern position.

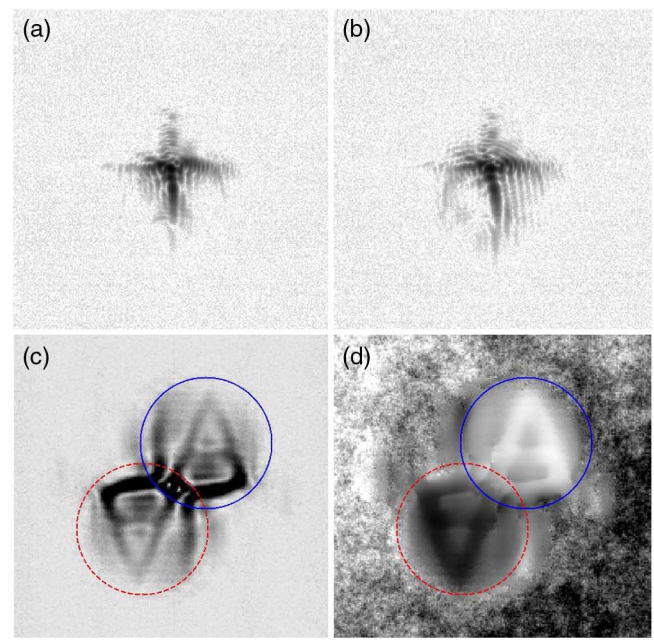

**Fig. 14** The PSFs and dOTF. The baseline (a) and modified (b) PSFs with a single actuator displaced by an additional 250 nm. The amplitude (c) and unwrapped phase (d) of the dOTF. The blue-solid circle indicates the approximate region of $\delta\mathcal{O}_+$ and the red-dashed circle shows $\delta\mathcal{O}_-$. The amplitude image shows a great deal of structure, some of which is bona fide amplitude variation and some that is the result of the $\delta\psi$ convolution. The phase image clearly shows the "A" pattern including influence function effects in the pattern relief.

The dOTF allows the use of any small pupil modification as a probe. The result is an estimate of the complex pupil field convolved with the introduced change in the pupil field. Deconvolution will probably be quite successful with dOTF, especially when extra assumptions are included, such as knowing that the pupil mask is binary or the aberrations are only phase. Then the amplitude of the dOTF will become an extremely sensitive measure of the phase variations. A segmented DM with piston-tip-tilt control can be used to create a Fourier component field change over the actuator used to modify the field. By adding several such modifications with more images, it may be possible to Fourier synthesize a very sophisticated convolution kernel, enabling measurement of field structures much smaller than the segment. The obscured overlap region in the dOTF can already be made quite small, but it can be eliminated altogether by taking another image with a different pupil modification. Similar interesting phenomena remain to be explored and exploited, such as using under-sampled PSFs and large bandwidths with more than one pupil change to achieve a complete map of the pupil field.

## 5 Summary

The dOTF method is a new and remarkably simple way to measure the complex amplitude of the pupil field in an optical system, including aberrations and transmission effects that occur between the pupil and the imaging sensor. The measurements can be made with the optical system in its working configuration, with very little or no impact on the system. We demonstrated the dOTF using only simple tools, sources and cameras, typically achieving a 5 nm rms accuracy in about 1 s. Better cameras and faster pupil modifications should allow construction of wavefront cameras at video frame rates or faster. We demonstrated the use of both transmission blockages and local phase changes to introduce the dOTF pupil change. The use of an extended pupil modification area blurs the result in a way that likely can be deconvolved, but also opens the possibility for field processing and detecting and measuring small-scale structures in the pupil field. Optical bandwidth introduces a proportional radial blurring centered on the pupil modification. While this method is most obviously applicable to optical test applications, dOTF will likely find uses in a large array of applications.

## Acknowledgments

This work was supported by the National Science Foundation Grant Nos. AST-0804586 and AST-0904839.

**Johanan L. Codona** received his BS degree in applied physics from the California Institute of Technology in 1980. He received his PhD in applied physics from the University of California, San Diego in 1985. His doctoral research involved the theory and application of Wave Propagation through Random Media (WPRM), with an emphasis on optical scintillation in the atmosphere and radio scintillation in astronomy. He was a Distinguished Member of Technical Staff (DMTS) at AT&T Bell Laboratories (later Lucent Technologies), where he worked on underwater acoustics and anti-submarine warfare. In 2002, he joined the Center for Astronomical Adaptive Optics (CAAO) at Steward Observatory, University of Arizona, where he has been a Senior Research Scientist. His research interests include adaptive optics, high-contrast imaging and instrumentation, coronagraphy, and wavefront sensing.